\documentclass[twoside,twocolumn]{article}
\usepackage{xr}
\usepackage[sc]{mathpazo} 
\usepackage[T1]{fontenc} 
\usepackage{lmodern} 
\usepackage{microtype} 

\usepackage[english]{babel} 
\usepackage{cite}
\usepackage[super,numbers,sort&compress]{natbib}

\usepackage{graphicx}
\usepackage{color}
\usepackage[top=2cm,bottom=2cm,left=1.5cm,right=1.5cm,columnsep=20pt]{geometry} 
    {\large\begin{scshape} Computational Details\end{scshape}%
    \par\medskip\normalsize}%
    {}%
    {\large\begin{scshape} Acknowledgement\end{scshape}%
    \par\medskip\normalsize}%
    {}%
    {\large\begin{scshape} Supporting Information\end{scshape}%
    \par\medskip\normalsize}%
    {}%
\usepackage{multirow} 
\usepackage[normal, small,labelfont=bf,up,textfont=it,up]{caption} 
\usepackage{booktabs} 

\usepackage{lettrine} 

\usepackage{enumitem} 
\setlist[itemize]{noitemsep} 

\usepackage{abstract} 

\usepackage{titlesec} 
\renewcommand\thesection{\Roman{section}} 
\renewcommand\thesubsection{\roman{subsection}} 
\titleformat{\section}[block]{\large\scshape\centering}{\thesection.}{1em}{} 
\titleformat{\subsection}[block]{\large}{\thesubsection.}{1em}{} 

\usepackage{fancyhdr} 
\usepackage{titling} 

\usepackage{hyperref} 

\pretitle{\begin{center}\Large\bfseries} 
\posttitle{\end{center}} 

\def\bea{\begin{eqnarray}}
\def\eea{\end{eqnarray}}
\def\ben{\begin{equation}}
\def\een{\end{equation}}
\def\benu{\begin{enumerate}}
\def\enu{\end{enumerate}}

\def\bei{\begin{itemize}}
\def\eei{\end{itemize}}
\def\beit{\begin{itemize}}
\def\eit{\end{itemize}}
\def\benu{\begin{enumerate}}
\def\enu{\end{enumerate}}

\def\sss{\scriptscriptstyle\rm}

\def\1var{(\bx_1...\bx\N)}

\def\bx{{x}}

\def\x{_{\sss X}}
\def\c{_{\sss C}}

\def\xc{_{\sss XC}}

\def\N{_{\sss N}}

\def\sph_int{ {\int d^3 r}}

\usepackage[table]{xcolor}
\usepackage{amsmath}
\usepackage{graphicx}

\def\d{_{\sss D}}
\def\f{_{\sss F}}

\title{Density sensitivity of empirical functionals}

\author{%
\textsc{Suhwan Song$^a$, Stefan Vuckovic$^b$, Eunji Sim$^{a,}$\thanks{esim@yonsei.ac.kr}, and Kieron Burke$^b$} \\ 
\normalsize $^a$Department of Chemistry, Yonsei University, 50 Yonsei-ro Seodaemun-gu, Seoul 03722, Korea \\ 
\normalsize $^b$Departments of Chemistry and of Physics, University of California, Irvine, CA 92697, USA \\  
}
\date{\today} 

\usepackage{xr}
\usepackage{float}

\renewcommand{\maketitlehookd}{%
\begin{abstract}
\noindent 
Empirical fitting of parameters in approximate density functionals
is common. Such fits conflate errors in the self-consistent
density with errors in the energy functional, but density-corrected
DFT (DC-DFT) separates these two.  We illustrate with
catastrophic failures
of a toy functional applied to $H_2^+$ at varying
bond lengths, where the standard
fitting procedure misses the exact functional; Grimme's D3 fit to noncovalent
interactions, which can be contaminated by large density errors
such as in the WATER27 and B30 datasets; and double-hybrids
trained on self-consistent densities, which can perform poorly on systems with
density-driven errors.
In these cases, more accurate results are
found at no additional cost, by using Hartree-Fock (HF)
densities instead of self-consistent densities.
For binding energies of small water clusters,  errors are greatly reduced.
Range-separated hybrids with 100\% HF at large distances suffer
much less from this effect.
\end{abstract}
}

\begin{document}

\maketitle


\sf

For the last quarter century, fitting of empirical parameters in 
approximate exchange-correlation functionals has been popular, especially
given the early successes of Becke88 exchange,\cite{B88} Lee-Yang-Parr
correlation,\cite{LYP88}
and the global hybrid ideas of Becke,\cite{B93} ultimately leading to the hugely
successful B3LYP.\cite{SDC94}   Since then, the number of functionals and the 
number of parameters has proliferated,\cite{YLT16,MH17} and often dozens
of parameters are fitted to dozens of databases, with thousands of benchmark
data.

There are many pitfalls to such fitting, but we focus on just 
one.  This danger is unambiguous, has nothing to do with choices of
parameters or datasets, and entirely avoidable.  Almost all such
fittings consist of running one or more self-consistent DFT calculations,
evaluating an energy difference, and comparing it with a (presumably accurate)
energy from the database.  (In the case of bond lengths, the difference
is an infinitesimal, determining where an energy derivative vanishes).
The accuracy of self-consistent densities was recently highlighted,\cite{MBSP17}
and how errors in the density can be related to errors
in the energy.\cite{CC90, OB94, JS08, BLGI09, KSB11, VPB12, KSB13, KSB14, KPSS15, WNJK17, SKSB18, SSB18, KSSB18}

{\em Background:}
The theory of density-corrected DFT (DC-DFT) has been developed over the past decade.\cite{KSB11}
Whenever a self-consistent (SC) DFT calculation is run, there are two
distinct sources of error. The total error of such calculations is
$\Delta E = \tilde E [\tilde n] - E[n]$, where $ E$ and $ n$ are the exact 
energy functional and density, and $\tilde E$ and $\tilde n$ 
are their approximate counterparts. We decompose $\Delta E$ as\cite{KSB13,SSB18,NSSB20}:
\ben
\Delta E=\underbrace{\tilde E[\tilde n] - \tilde E[n]}_{\Delta E\d} 
+
 \underbrace{ \tilde E[n] - E[n]}_{\Delta E\f}.
\label{eq:FD}
\een
where $\Delta E\f$ is the functional error, defined
as the error that would be found if the exact density were used, while
$\Delta E\d$ is the (usually much smaller)
contribution to the energy error due to the error in the self-consistent
density.  

So long as density-driven errors were small compared to the functional
errors (as was the case in the halcyon days of B3LYP), they were irrelevant.
But in the modern era of vast databases that include weak interactions, stretched bonds,
etc., these errors are sometimes as big as (or larger than) the 
functional errors.\cite{WNJK17, KSSB18}  
However, the common practice of direct comparison
with accurate energies
conflates both errors and cannot distinguish the two. Recent advances
in machine learning of density functionals target the density as well
as the energy, and likely succeed because both errors are simultaneously
minimized.\cite{NAS19}

The cure for this difficulty is simple:  where relevant, empirical schemes
should be trained on purely functional errors, i.e., the functional
error of a parameterized 
approximation to the energy should be optimized against accurate energy databases,
rather than the self-consistent error.  For calculations that are not density-sensitive,
the differences are so small as to make this irrelevant.
But for those that are, this procedure isolates the self-consistency
error and so avoids the corruption of the optimization process, allowing
density-sensitive cases to be included even in training.

The current paper highlights the consequences of ignoring this distinction when
optimizing parameters in empirical functionals.   We first create a
totally artificial problem to emphasize the difficulties, especially
when one uses a semilocal approximation for the self-consistent density
but a more accurate form for the energy.  In this case, we show how
the exact functional is missed by the standard procedure.   Next, we take
the D3 correction of Grimme and co-workers,\cite{GAEK10} and show how, if complexes with
large density-driven errors are naively included, the results become
noticeably worse.  On the other hand, the use of DC-DFT allows previous good
results to be retained, and the more difficult complexes to be included.
We also apply our method to double-hybrids (DHs), 
producing a combination that competes with similar functionals, 
but still works when the density sensitivity is large. 
Finally, we find that empirical range-separated hybrid functionals
suffer less from  density-driven errors than their conventional
global counterparts.

For the purposes of this paper, we write a
4-parameter double-hybrid functional ($DH4p$) as:
\ben
\begin{split}
E\xc^{\rm DH4p}&=E\x^{\sss Slater}+ \alpha (E\x^{\sss HF}-E\x^{\sss Slater}) + \beta (\tilde E\x^{\sss GGA}-E\x^{\sss Slater}) \\
&+\gamma \tilde E\c^{\sss GGA}+\delta E\c^{\sss ab-initio},
\label{eq:dh4p}
\end{split}
\een
where $E_X^{Slater}$ is the local density approximation for exchange,
$E\x^{HF}$ is the HF exchange, $\tilde E\x^{GGA}$ and $\tilde E\c^{GGA}$
denote the approximate GGA exchange and correlation energy, respectively, and
$E\c^{ab-initio}$ is the correlation energy
from an ${\it ab-initio}$ calculation such as MP2.
The standard procedure then is to run self-consistent calculations
of Eq.~\ref{eq:dh4p} without the ${\it ab-initio}$ correlation, but
evaluate
energies with the full DH expression on the orbitals.\cite{G06,STS11,MS19}
The parameters are then chosen to minimize errors 
for specific molecular datasets.
  As we show,
this assumes
that density-driven differences between this and doing the entire procedure
self-consistently are negligible.   

Often, highly-accurate densities required in Eq. 1  are too expensive to calculate.
A practical measure of density sensitivity 
is given by:\cite{SSB18,KSSB18,NSSB20}
\ben
\tilde S= \left | \tilde E [n_{\sss LDA}]-\tilde E [n_{\sss HF}] \right|,
\label{eq:s}
\een
where tilde indicates a given functional approximation.  Given the HF tendency
to overlocalize, and the LDA tendency to delocalize, and that both are non-empirical,
$\tilde S$ is a practical guide to the density sensitivity of a given reaction and
approximate functional.  For small molecules, $\tilde S > $ 2 kcal/mol implies
density sensitivity and suggests DC-DFT will improve a functional's performance~\cite{SSB18}.
In such cases, usually the HF density is sufficient to produce improved energies (HF-DFT).

\begin{figure}[htb]
\centering
\includegraphics[width=0.95\columnwidth]{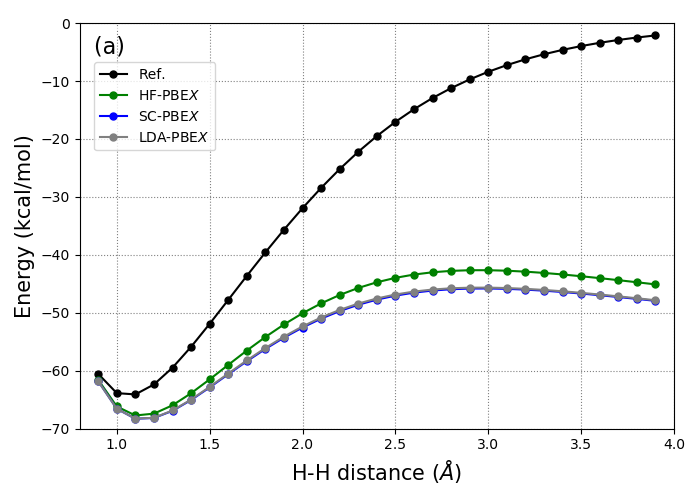}
\includegraphics[width=0.95\columnwidth]{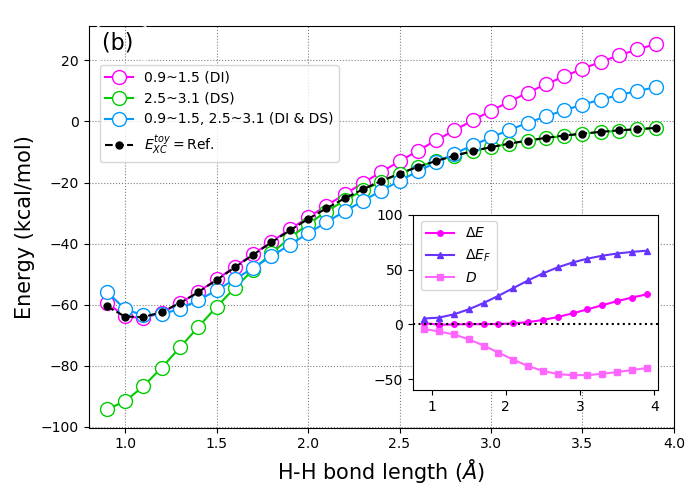}
\caption{Potential energy surface (PES) of H$_{\sss 2}^{\sss +}$ from:
(a) exactly (black), self-consistent PBE$X$ (blue) and PBE$X$ on the exact (HF) density (green) and on the LDA
density (grey);
(b) the toy functional of Eq.~\ref{eq:dh4p} with $\gamma=\delta=0$ and no HF in the self-consistent
density, with the a and b parameters optimized in different regions:
(magenta) the density-insensitive (DI) region  (0.9\AA-1.5\AA),
(green) the density-sensitive (DS) region (2.5\AA-3.1\AA),
(blue) combination of both DS and DI regions.
The inset shows 
$\Delta E$ decomposition for the toy functional trained on
the DI region.
See also Fig.~\ref{v2-fgr:divide}, \ref{v2-fgr:h2+pes}, and Table~\ref{v2-tbl:h2+para1}.
}
\label{fgr:h2+}
\end{figure}

{\it Illustration: Missing the exact solution for one electron--} In this section, we illustrate
the dangers of ignoring the distinction between
density-driven and functional errors in a simple, toy model:  A simplified hybrid
applied to the elementary case of H$_2^+$ as a function of bond length, which is
a paradigm of self-interaction error, or
more generally, delocalization error.\cite{CMY08,CMY11}  Standard semilocal approximations
yield long-recognized catastrophic errors as the bond is stretched, missing
entirely the dissociation limit (see Fig~\ref{fgr:h2+}).\cite{CMY11}
A HF calculation trivially gets
this exactly right, since it is exact for (fully spin-polarized) one-electron systems.

Fig.~\ref{fgr:h2+}(a) shows the exact binding curve (black)
easily found by HF, and two other curves of the PBE$X$ evaluated
either self-consistently (blue) or on the HF density (green).
The largely irrelevant difference between blue and green curves show that this
is a true functional error, not a density-driven one.  Even on the exact
density, PBE$X$ fails very badly as the bond is stretched.   However, the
difference in the two curves becomes greater than 2 kcal/mol at about $1.5 \AA$, showing 
a density sensitivity (the curve with LDA density is indistinguishable from the
self-consistent curve) in this problem.  (Standard HF-DFT
produces accurate curves for heteronuclear diatomics, not
homonuclear ones.\cite{KPSS15,WNJK17})

Now, to mimic the standard DH procedure, we perform self-consistent calculations without
the HF contribution (since it yields the exact answer in this case), but evaluate
the energy with it included.
We apply the DH philosophy to our H$_2^+$ molecule, using different separations
to generate datasets.  Because this is a one-electron system, we simplify the general
DH form to just exchange, setting $\gamma = \delta = 0$ in Eq.~\ref{eq:dh4p}, and
use the PBE exchange~\cite{PBE96} as a GGA.
Fig.~\ref{fgr:h2+}(b) shows the
results of training in the density-sensitive (stretched, DS)
and density-insensitive (near equilibrium, DI) region of the binding curve. 
In each case, the optimal parameterization
yields accurate energies on the training data, but fails badly outside the 
training range.
Even a combination of both equilibrium and stretched data does not help much. 
 
How can this be happening?  Obviously, if we set $\alpha=1$ and $\beta=0$ in Eq.~\ref{eq:dh4p}, we
get HF, and so produce the exact answer.  But, because the self-consistent calculation
uses only a GGA form, which has an unbalanced self-interaction error as the bond
is stretched, the exact result is never found.  To quantify, we define
\ben
D[n']=\tilde E[n'] - \tilde E[n],
\label{eq:DF}
\een
generalizing\cite{VSKS19} $\Delta E\d$ to arbitrary densities ($D[\tilde n]=\Delta E\d$, and $D[n]=0$).
We decompose the error for the functional trained near equilibrium, 
showing $\Delta E\f$ and $D$ in the
inset of Fig.~\ref{fgr:h2+}(b). The optimal parameters (which are nonsensical, see Table~\ref{v2-tbl:h2+para1} of the supporting information) keep the total
error to a minimum in the training region where
$\Delta E\f$ and $D$ cancel each other by being about equal and opposite.  
Outside the training
region of our H$_2^+$ curve, this artificial cancellation of errors fails badly.
Obviously, we trivially solve this toy problem if we always train on the HF (exact, in this case) density instead
of the self-consistent GGA density.

\begin{table}[htb]
\centering
\begin{tabular}{lcccc}
\hline
            & \multicolumn{2}{c}{[SC]} & \multicolumn{2}{c}{[HF]} \\
opt. dataset            & DI   & DS   & DI   & DS   \\
\hline
without opt.	&	1.53 	&	2.90 	&	1.89 	&	4.95 	\\
D3$_{\sss orig}$	&	0.43 	&	6.74 	&	0.42 	&	1.20 	\\
12DB	&	0.48 	&	5.66 	&	0.31 	&	0.98 	\\
DS-12DB	&	1.47 	&	2.96 	&	0.38 	&	0.87 	\\
DI-12DB	&	0.42 	&	6.53 	&	0.31 	&	1.01 	\\
\hline
\hline
\end{tabular}
\caption{
Mean absolute errors (kcal/mol) of PBE and modifications on density-insensitive (DI) and
density-sensitive (DS) test cases (columns) versus optimization on various databases
(rows), with self-consistent (SC) densities on left and HF densities on right.
D3$_{\sss orig}$ denotes the original Grimme dataset, 12DB is our large (320 values) mixed dataset,
DI-12DB are its 274 DI cases, and DS-12DB its 46 DS cases.}
\label{tbl:optmae}
\end{table}

{\it DFT-D3 for weak interactions--} 
The D3 empirical correction of Grimme and co-workers has become a standard technique
for improving 
the accuracy of DFT approximations when applied to noncovalent interactions.\cite{GAEK10, CEHN19}
While most such calculations are density insensitive, 
DFT calculations of
specific types of noncovalent interactions, such as halogen bonds,
are plagued by density errors, which can
be larger than the D3 correction itself.\cite{KSSB18} 

\begin{figure*}[htb]
\centering
\includegraphics[width=1.9\columnwidth]{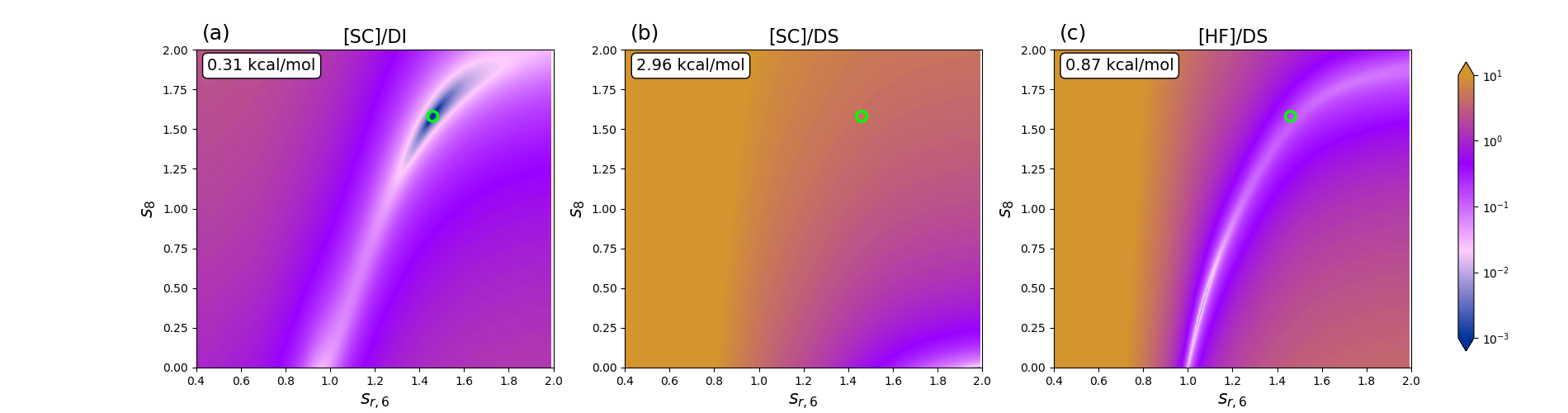}
\caption{
Mean absolute error (MAE) of PBE-D3 as a function of dispersion parameters\cite{GAEK10},
for various
densities and test sets: (a) self-consistent (SC) density on density-insensitive (DI) cases, 
(b) SC density for density-sensitive (DS) cases, and
(c) HF density for DS cases.
Contours are shifted by the minimum value (upper left corner)
for clarity.
The green circle is at the position of the global minimum of the panel~(a).
}
\label{fgr:surf}
\end{figure*}

HF-DFT, as a simple form of DC-DFT,  fixes this problem by replacing the
SC density with the HF density, on which semilocal functionals yield more
accurate energies in such cases.\cite{KSB13,SSB18,KSSB18,VSKS19}
It was recently shown that the use of the HF density in place of the exact density
introduces much smaller errors than the improvements made by HF-DFT.\cite{NSSB20}
(As an aside, this does not imply
that 
the pointwise accuracy of the underlying HF density is better than 
that of SC-DFT densities.\cite{SSB18})

The example of Ref.~\citenum{KSSB18} was an extreme case.
Here we study the effects of density sensitivity on SC-DFT-D3 calculations of weak interactions
when they are more subtle.
We use 12 datasets (7 from the original D3 parameterization~\cite{GAEK10})
of noncovalent interactions (320 data points in total, see Table~\ref{v2-tbl:dbdes} of the supporting information).\cite{GHBE17} 
The data points are classified as DS or DI based on their PBE
sensitivity, $S^{\rm PBE}$ (see Eq.~\ref{eq:s} and Fig.~\ref{v2-fgr:s1}).
Only 46 are DS, and these are mostly from 
B30\cite{BAFE13} and WATER27\cite{GHBE17}, with only one 
such data point present in the dataset used for the training of
the original D3 parameters.

In Table~\ref{tbl:optmae}, 
we demonstrate the importance of accounting for the density sensitivity when
optimizing parameters for D3 corrections.  The first two numbers in the 2nd column show the
dramatic reduction in error in the PBE functional when the original D3 correction is made,
on the density-insensitive cases.
The next entry shows that when we optimize over  our much expanded database, the errors for
DI cases are only slightly worse.  But if we optimize specifically over our DS cases (4th entry),
this greatly
worsens results on our DI test cases.

Moving over one column, we find results when tested on the DS cases.  Now the original
D3 parameterization yields a large (greater than 6 kcal/mol) error, demonstrating that
density-sensitivity creates large errors.  Even when optimized for DS cases, the error
remains about 3 kcal/mol.  

In the next column, we report the DI test results, but using HF densities instead of SC
densities.  In all cases of interest, the errors
are slightly reduced once D3 with any of the parameters is turned on.
The errors fall by more than
a factor of 6 if the D3 is trained on the DI cases.
Furthermore, the differences between the optimal D3 parameters for DS and DI cases 
are much smaller when HF densities are used.
Fig.~\ref{fgr:surf} shows the variation of the error with parameters.
Fig.~\ref{fgr:surf}(a) shows the usual case (SC densities on DI cases).  Fig.~\ref{fgr:surf}(b)
is SC densities on DS cases, showing a totally different landscape.
A green circle lying at the minimum of the case (a) is denoted in all three panels.
Fig.~\ref{fgr:surf}(c)
is HF densities on DS cases, showing about the same landscape as (a).  

Finally,
the fourth column of Table~\ref{tbl:optmae} shows results on the DS cases using HF densities.
While overall, these are much less
accurate than the DI cases (by about a factor of 3), they are much better than 
those of column 2, which uses SC densities 
 
From these findings we can also see the effects of including DS cases in the training set. Their naive inclusion without the density correction via HF-DFT gives some improvements for DS cases at the cost of deteriorated accuracy for DI cases resulting from the abrupt changes in the optimal parameters.  On the other hand, after the density correction is applied, the inclusion of DS cases in the training set improves their accuracy without the side effects for DI cases (Table~\ref{tbl:optmae}) and without abrupt changes in the parameter landscape (Fig.~\ref{fgr:surf}).

\begin{figure}[htb]
\centering
\includegraphics[width=0.95\columnwidth]{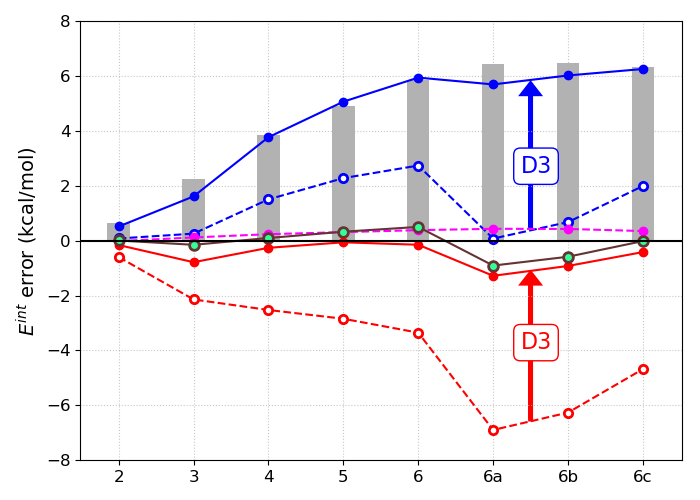}
\caption{
PBE binding energy error for small water clusters, 
$E^{\sss int}=nE_{\sss H_2 O}-E_{\sss (H_2 O)_n}$ ($n=2\sim6$), in WATER27 dataset.
Blue denotes self-consistent (PBE), while
red is for the HF density (HF-PBE); dashed is
without dispersion correction, while solid denotes with D3 (revised
is similar to original).
The gray bar shows the density-sensitivity of Eq.~\ref{eq:s}. 
For comparison, we also show 
$\omega$B97M-V (magenta) 
and BL1p (green, defined later in text) results.}
\label{fgr:newwater}
\end{figure}

\begin{figure}[!h]
\centering
\includegraphics[width=1.0\columnwidth]{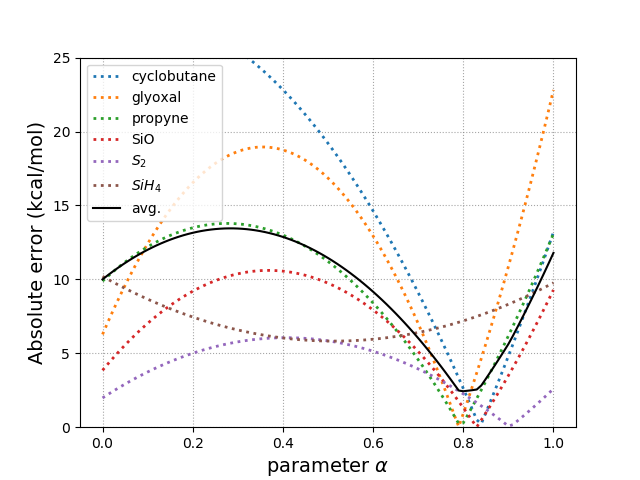}
\caption{
Absolute errors for the AE6 dataset of BL1p as a function of $\alpha$
(see Eq.~\ref{eq:dh1p} ) for individual molecules (dashed lines). In the black solid curve, where the averaged errors are shown, the minimum is achieved at $\alpha=0.82$.}
\label{fgr:dalpha}
\end{figure}

Most of the DS noncovalent complexes used in the training set in Table~\ref{tbl:optmae}
belong to the B30 and WATER27 datasets. 
In Fig.~\ref{fgr:newwater}, we 
compare errors of SC-PBE and HF-PBE, with and without 
the (revised)D3 correction, for binding energies of small water clusters
of the WATER27 dataset. The standard
DFT calculations of these binding energies are
highly DS, as shown by the
large values for $S^{\rm PBE}$ shown in Fig.~\ref{fgr:newwater}. 

We see that HF-DFT corrections are larger than D3 here, and that D3 on self-consistent densities actually
corrects in the wrong direction. 
HF-DFT-D3 reduces errors for the largest clusters from about 6 kcal/mol
to less than 1 kcal/mol, and thus delivers the performance comparable to $\omega$B97M-V~\cite{MH16}, which includes 
nonlocal correlation~\cite{VV10}, and BL1p (a DH that will be introduced later).

\begin{figure*}[!h]
\centering
\includegraphics[width=2\columnwidth]{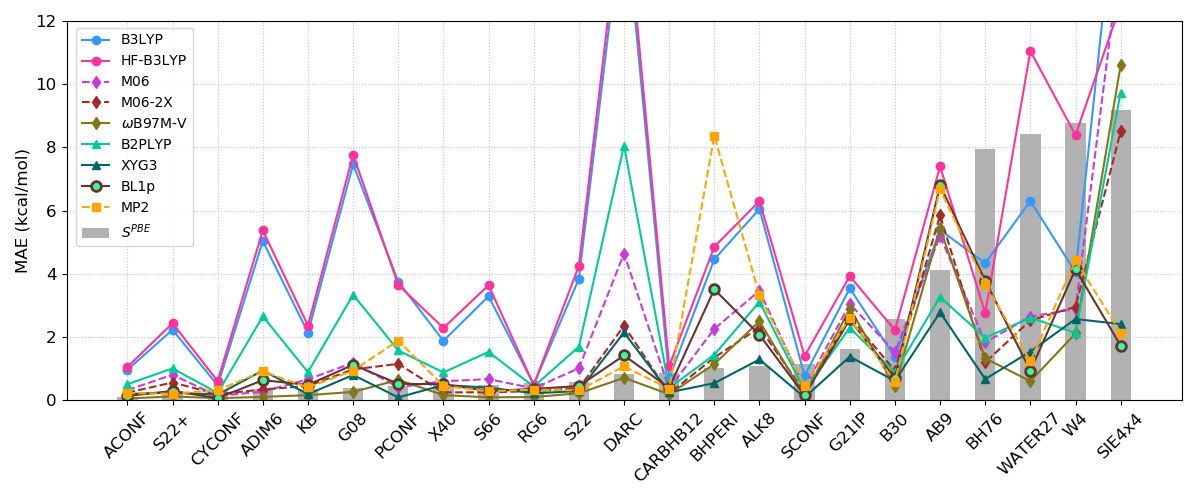}
\caption{
MAEs for several methods on many databases: BL1p,  other double-hybrids (B2PLYP and XYG3), 
hybrids (B3LYP,M06, M06-2X),
range-separated meta-GGA hybrid ($\omega$B97M-V), and MP2. }
\label{fgr:1p}
\end{figure*}


{\it Double-hybrids--}  
The energy functional  of widely popular DHs (e.g. Eq.~\ref{eq:dh4p}) is typically
evaluated on the hybrid density and orbitals found in a self-consistent
calculation that neglects the $E_C^{ab-initio}$ term.\cite{G06,STS11}
In contrast, we find that HF-DHs obtained by applying a
DH energy expression to the HF density and orbitals yield
an overall accuracy competitive with their standard counterparts,
but
remain accurate for cases where the standard DHs fail due to density sensitivity.
We test the HF-DH idea with only one empirical parameter:\cite{STS11}
\ben
\begin{split}
E\xc^{\rm DH1p}&=E\xc^{\sss GGA} + \alpha(E\x^{\sss HF}-E\x^{\sss GGA})\\
&+\alpha^2 (E\c^{\sss ab-initio} - E\c^{\sss GGA})
\end{split}
\label{eq:dh1p}
\een
as suggested by Sharkas {\it et al.} based on {\em adiabatic connection} arguments.\cite{STS11}
(This is Eq.~\ref{eq:dh4p} where $\beta=1-\alpha$ and $\delta=1-\gamma=\alpha^2$.)
To construct a HF-DHs based on Eq.~\ref{eq:dh1p}, we use here a combination of:
B88 exchange\cite{B88}, semilocal LYP correlation;\cite{LYP88} and  
MP2 correlation for $E\c^{\sss ab-initio}$\cite{MolPle-PR-34}. 
We call this functional BL1p.
Also, see Fig.~\ref{v2-fgr:ae6} to compare 1DH-BLYP (BL1p[SC]) of Ref.~\citenum{STS11} and BL1p[HF].
Here we do not aim at reaching the accuracy limit of the HF-DH approach. This is already prohibited by a functional form of Eq.~\ref{eq:dh1p}, which contains only one empirical parameter. 
Our goal is to show that this approach delivers an overall performance comparable
to the standard DHs while not being plagued by large density-driven errors.
Thus, we perform the optimization of $\alpha$ of Eq.~\ref{eq:dh1p} in an old-fashioned way, by training BL1p[HF] on the AE6 dataset, containing atomization energies of 6 molecules.~\cite{LT03}
The results of the training are shown in Fig.~\ref{fgr:dalpha}. At $\alpha=0$, our BL1p reduces to HF-BLYP, whereas at $\alpha=1$,
it reduces to MP2.
The optimal BL1p that minimizes MAE for AE6 has $\alpha=0.82$, which varies little between
molecules, except for 
SiH$_4$ whose minimum is much shallower.
Also, the MAE of optimal BL1p is about 7.5 kcal/mol smaller than 
the $\alpha=0$ case (HF-BLYP) and about 9 kcal/mol than the 
$\alpha=1$ case (MP2).

In Fig.~\ref{fgr:1p}, we compare the performance of BL1P with
the standard DHs (B2PLYP\cite{G06} and XYG3\cite{ZXG09}), hybrids (B3LYP, M06, M06-2X), 
and also with the range-separated functional ($\omega$B97M-V\cite{MH16}), which we detail in the supporting information. 
This figure shows that the one-parameter BL1p, trained only 6 atomization energies,
yields an accuracy that is competitive with the
standard DHs for all databases, and works for noncovalent interactions,
without using Grimme's empirical correction.
Usually, we recommend against using the
HF density when it suffers from spin-contamination.\cite{KSB13}
Nevertheless,
for all data in this section, we include the spin-contaminated cases
for fair comparison. Performance without spin-contaminated cases is shown in the supporting information (see Table~\ref{v2-tbl:mae}).

\begin{figure}[!htb]
\centering
\includegraphics[width=0.95\columnwidth]{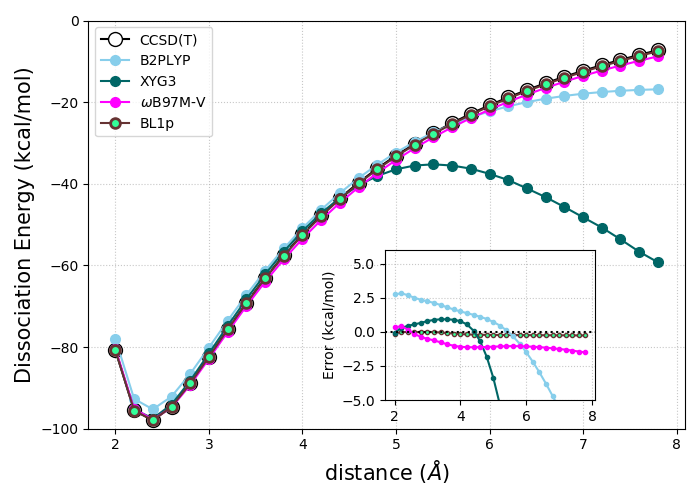}
\caption{
Dissociation curves of NaCl obtained from various approaches. 
For stretched bond lengths, 
standard 
double-hybrid functionals fail due to the density-driven errors (see Ref.~\citenum{KPSS15}).
}
\label{fgr:nacl}
\end{figure}

\begin{figure}[!htb]
\centering
\includegraphics[width=0.95\columnwidth]{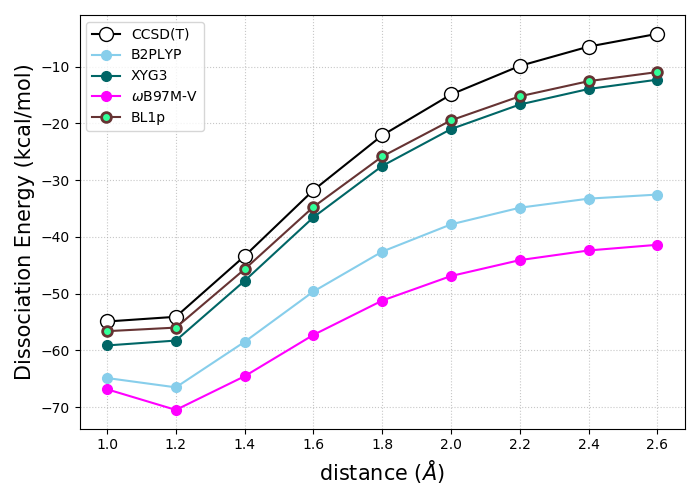}
\caption{
Dissociation curve of He$_2^+$ obtained from various functionals. 
See also Fig.~\ref{v2-fgr:he2+inv}, showing that the errors of standard functionals for He$_2^+$ are
mostly functional errors, since self-consistent results are almost
identical to those when the functionals are applied to 
accurate densities (obtained from the Kohn-Sham inversion scheme from the CCSD wavefunction~\cite{NSSB20}).}
\label{fgr:he2+}
\end{figure}

Returning to our starting point, stretched NaCl is a prototypical case where
self-consistent hybrids and GGAs are contaminated
by large density errors.\cite{KPSS15} These errors are typical of semilocal functionals for
dissociating heterodimers.\cite{GriBae-PRA-96,GiaVucGor-JCTC-18}
HF densities fix this problem, and HF-DFT is able to dissociate heterodimers correctly.\cite{KPSS15}
From Fig.~\ref{fgr:nacl}, in contrast to a standard DHs (B2PLYP and XYG3 shown here) that fail at large bond lengths,
our BL1p, as a representative of HF-DH, dissociates NaCl correctly (See also Fig~\ref{v2-fgr:xyg3nacl}). 

Another case where BL1p outperforms other methods is the SIE4x4 dataset, 
containing four positively charged dimers at four different separations, 
where standard DFT methods have large self-interaction error.~\cite{GHBE17}
Fig.~\ref{fgr:he2+} shows the dissociation curve of He$_2^+$, as
a representative of this dataset. First, the errors of the standard DFT
methods for He$_2^+$ are almost entirely functional errors (see Fig.~\ref{v2-fgr:he2+inv}), because
they differ little between accurate and self-consistent densities.
The accurate densities are obtained by Kohn-Sham inversion from 
CCSD densities.~\cite{NSSB20} In this way, the source of error of
the standard DFT for He$_2^+$ is very different from that of stretched NaCl.
Fig.~\ref{fgr:he2+} shows that, even though these are not density-driven
errors, the error of BL1p for He$_2^+$ is
much smaller than that of other approaches.

\begin{figure}[!htb]
\centering
\includegraphics[width=0.95\columnwidth]{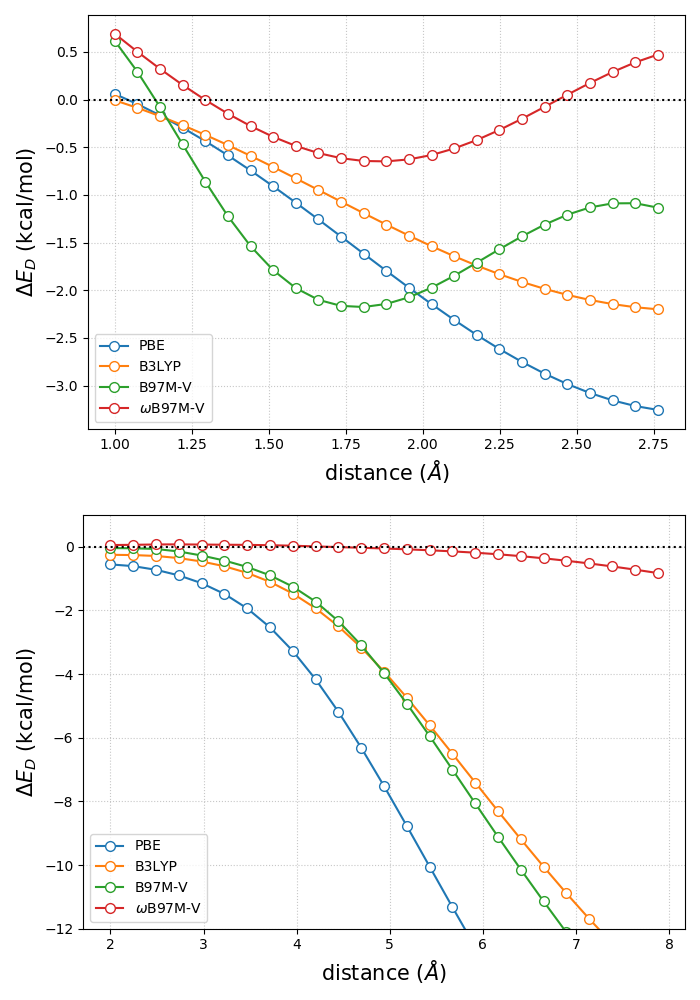}
\caption{
Density-driven errors (see Eq.~\ref{eq:FD}) of selected functionals along the dissociation curves of:e
H$_2^+$ (top panel) and NaCl (bottom panel). For H$_2^+$, the (exact) HF density is used to extract the density-driven errors. For NaCl, 
we use CCSD as a reference in tandem with the Kohn--Sham inversion scheme described in Ref.~\citenum{NSSB20} to obtain the 'exact' density and orbitals needed to isolate density-driven errors.}
\label{fgr:dde_fig}
\end{figure}

\begin{figure}[!htb]
\centering
\includegraphics[width=0.95\columnwidth]{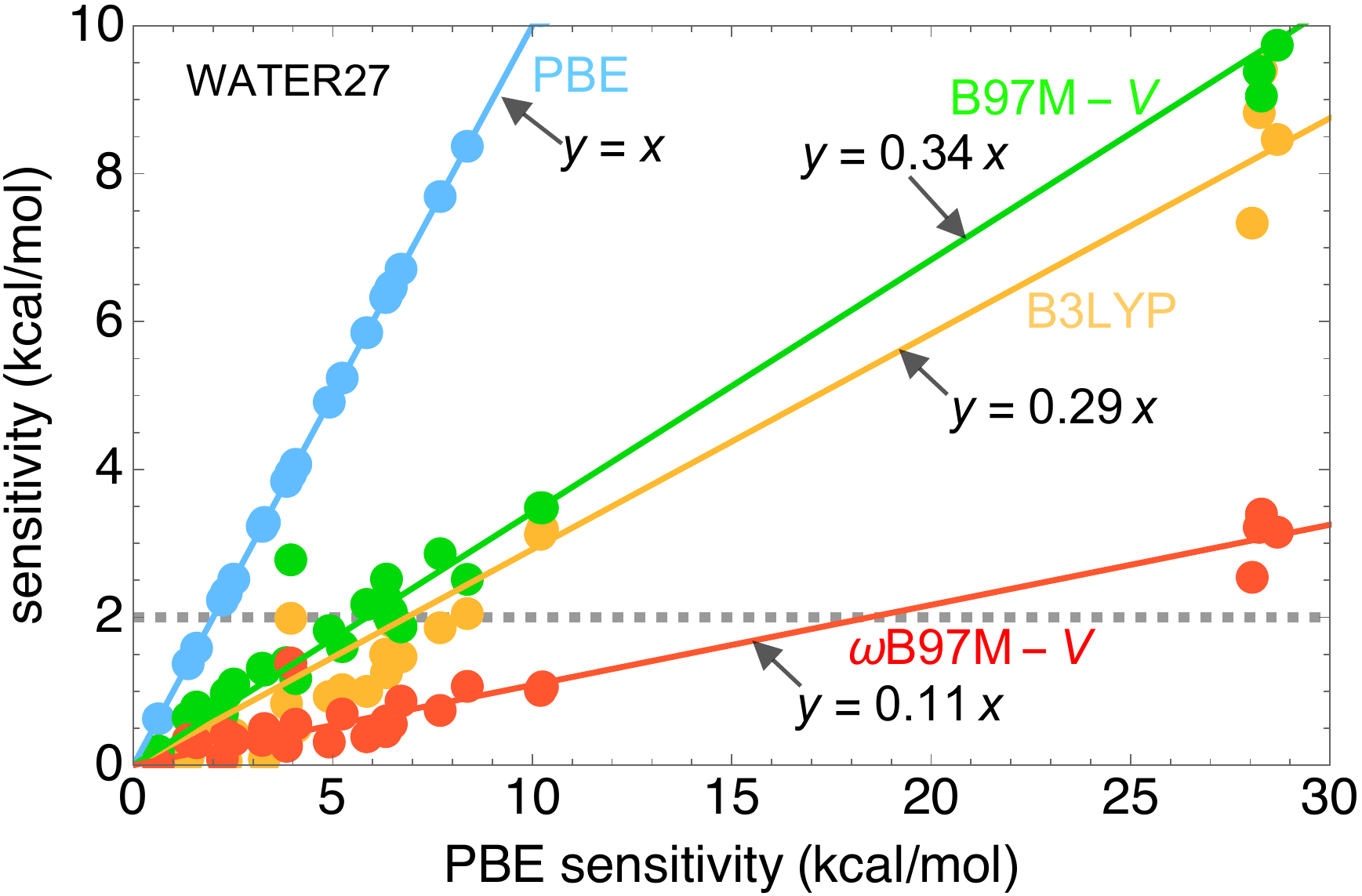}
\caption{
The sensitivity (see Eq.~\ref{eq:s}) of selected functionals vs. PBE sensitivities for binding energies of the WATER27 clusters.}
\label{fgr:w27}
\end{figure}

{\it Range-separated vs. conventional hybrids--} We have shown a number of examples
where large density-driven errors of conventional (global) hybrid functionals
are substantially reduced when they are evaluated on the HF instead of SC densities. 
Range-separated hybrids (RSHs) often use 100$\%$ of the HF exchange in the
long-range (lr)~\cite{VS06,MH14,MH16}, and so should suffer less from density-driven errors.~\cite{JOD13}.
To test this, we use $\omega$B97M-V as a representative of RSHs~\cite{MH16}, given
its remarkable performance for many of the databases in Fig.~\ref{fgr:1p}. 
We will compare $\omega$B97M-V with B97M-V, its conventional analog.~\cite{MH15}.
The density-driven errors of $\omega$B97M-V and B97M-V are shown in Fig.~\ref{fgr:dde_fig}
for our two standard cases, with PBE and B3LYP also shown for comparison. 
For H$_2^+$, the HF density is exact, while for NaCl,
we invert the accurate Kohn-Sham density from CCSD.~\cite{NSSB20}
In each case, the
density-driven error of $\omega$B97M-V is 
much smaller than that of the other functionals. 
It does not vanish, because of the semilocal part of the functional. 
We see similar behavior for larger systems where the error of conventional 
hybrid functionals is contaminated by the densities, and is much smaller in 
$\omega$B97M-V. Sensitivity plots are used as a diagnostic tool for density-driven errors,
and in Fig.~\ref{fgr:w27} we show that the 
sensitivity of $\omega$B97M-V for the WATER27 complexes is a fraction of that of B3LYP and B97M-V. 

{\it Concluding remarks--} We have shown the dangers 
of ignoring density errors in 
the construction of empirical approximations.
In our simple  H$_2^+$ example, a parameterized semilocal functional
trained on a limited region of the H$_2^+$
binding curve fails in all other regions.
Even high accuracy in the training region results from an enforced
error cancellation between 
the density and functional error (Eq.~\ref{eq:FD}), which fails outside this region.
We found that 
the 
standard DFT with empirical D3 corrections breaks down
in {\em density-sensitive} calculations of noncovalent systems, but is fixed by using the HF density. 

We also found that resilience to density-driven errors could be achieved with simple
1-parameter double-hybrids, once they are trained and applied to HF densities.
As always, our use of HF densities does not imply that
they are point-wise more accurate than self-consistent densities, but
simply that they yield more accurate energetics when a reaction is density sensitive.
Our BL1p is trained only on atomization energies of only 6 molecules, but its accuracy is  comparable to the standard doubled hybrids tested here. 
Moreover, $\omega$B97M-V outperforms BL1p for most of the datasets considered in Fig 5, except for the SIE4x4 dataset, where BL1p does much better. 
BL1p would also be beaten by $\omega$B97(2), a very recent highly accurate DH designed to improve over $\omega$B97M-V.~\cite{MHG18}
Given its excellent performance~\cite{MHG18,SSM19}, we expect it to beat BL1p on most of the datasets, but not SIE4x4.

Our goal here is not the introduction of a new empirical XC functional, 
but to illustrate contamination due to density errors in fitting procedures
and to show how minimizing the functional error can
improve the performance of empirical functionals. Thus, our primitively optimized BL1p
does not reach the accuracy limit of the HF-DH class of functionals. Technical advances in optimization and larger parameter spaces
could further improve its accuracy. Furthermore, to improve HF-DHs, one may also
use the new insights into functionals that explicitly depend on the
HF density obtained from the adiabatic connection that has the MP2 theory 
as its weak-interaction expansion.~\cite{SGVFG18,DGVMKSGG20}
Finally, we have found that using 
100$\%$ of HF exchange in range-separated hybrids means they suffer much less from
density-driven errors than their conventional counterparts. 

In summary, DFT energy errors can be separated into functional and density-driven using DC-DFT.
To avoid inaccuracies, empirical functionals can be trained on functional errors only, where practical. In cases of large density sensitivity, HF densities (unless flawed by, e.g., spin-contamination) are typically more useful than self-consistent semilocal densities.  With 100$\%$ exchange at large distances, range separated functionals are relatively density insensitive, and suffer much less from these issues.


\section*{Computational Details}
All HF, DFT, HF-DFT, and MP2 calculations have been performed with the TURBOMOLE v7.0.2.\cite{T15} and PYSCF v1.7.2.\cite{PYSCF} 
The following functionals have been used in DFT and HF-DFT calculations:
LDA (SVWN\cite{D30,VWN80}), GGA (PBE\cite{PBE96}, BLYP\cite{B88,LYP88}), 
mGGA (TPSS\cite{TPSS03}), hybrids (B3LYP\cite{SDC94}, PBE0\cite{BEP97}, M06, M06-2X\cite{ZT08}, B97M-V\cite{MH15}, $\omega$B97M-V\cite{MH16}, 
B2PLYP\cite{G06}, and XYG3\cite{ZXG09}). 
The scripts for performing HF-DFT energy calculations are available.\cite{TCCL}
Unless otherwise stated, the def2-QZVPPD basis set has been used.
All geometries
and the multiplicities except for the 
AE6~\cite{LT03} have been taken from Ref.~\citenum{GHBE17}.
Further computational details can be found in the supporting information. 



\section*{Supplementary Information}
\noindent
$\bullet$ Dataset description \\
$\bullet$ Optimized parameters and mean absolute error for $H_2^+$\\
$\bullet$ Mean absolute error value of Fig.~\ref{fgr:1p}\\

\section*{acknowledgement}
This work at Yonsei University was supported by the grant from the Korean Research Foundation (NRF-2020R1A2C2007468 and NRF-2020R1A4A1017737).
KB acknowledges funding from NSF (CHEM 1856165). 
SV acknowledges funding from the Rubicon project (019.181EN.026), which is financed by the Netherlands Organisation for Scientific Research (NWO).
We thank Professor Martin Head-Gordon and Dr. Narbe Mardirossian for stimulating discussions.





\bibliographystyle{unsrt}

\label{page:end}
\end{document}


\beginsupplement

\maketitle


\sf

\listoftables
\listoffigures
\clearpage



\subsection{Computational Details}

All HF, DFT, HF-DFT, and MP2 calculations have been performed with the TURBOMOLE v7.0.2.\cite{T15} and PYSCF v1.7.2.\cite{PYSCF} 
The following functionals have been used in DFT and HF-DFT calculations:
LDA (SVWN\cite{D30,VWN80}), GGA (PBE\cite{PBE96}, BLYP\cite{B88,LYP88}), 
mGGA (TPSS\cite{TPSS03}), hybrids (B3LYP\cite{SDC94}, PBE0\cite{BEP97}, M06, M06-2X\cite{ZT08}, B97M-V\cite{MH15}, $\omega$B97M-V\cite{MH16}, 
B2PLYP\cite{G06}, and XYG3\cite{ZXG09}). 
The scripts for performing HF-DFT energy calculations are available.\cite{TCCL}
Unless otherwise stated, the def2-QZVPPD basis set has been used.
For accelerating the self-consistent field (SCF) procedure,
we adopted the resolution of the identity approximation (RI-J) with def2-QZVPPD auxiliary basis set
for G08, PCONF, SCONF, and WATER27 dataset.
For the KB dataset, we omit the noble gas and the S22 dataset from the original KB65.
The energy and one-electron density convergence threshold have been set to 1e-8 and 1e-6 a.u., respectively. 
Numerical quadrature grids of size 4 have been used (grid size 4 in TURBOMOLE).
For VV10 correlation, 99 radial shells with 590 angular grid points per shell are used
with the SG1 prune.
For all open shell calculation, unrestricted scheme is used.
The parameters of Table~\ref{tbl:h2+para1} are first found in the global optimizer {\em shgo} and then optimized locally with the {\em Nelder-Mead} method in the SCIPY package with the 1e-8 convergence criterion.
For revised D3, we scanned $0.00<s_{r,6}<2.00$ and $0.00<s_{8}<2.00$ (0.01 grid spacing)
for all 320 datapoints  in 12 databases (DB) 
(marked by an asterisks in table~\ref{tbl:dbdes}) 
for all XC functionals except $\omega$B97M-V, XYG3 and HF-DHs.
The $\alpha$ value for HF-DHs is optimized with the {\em slsqp} method in the SCIPY package.
All geometries
and the multiplicities except for the AE6~\cite{LT03} have been taken from Ref.~\cite{GHBE17}.

\begin{table}[]	
\footnotesize
\centering				
\resizebox{\columnwidth}{!}{\begin{tabular}{lrl}				
\hline					
DB	&	no. points	&	Description	\\
\hline					
RG6*	&	6	&	rare gases	\\
ACONF*	&	15	&	alkane conformers	\\
S22+*	&	66	&	non-covalent interaction	\\
CYCONF*	&	10	&	cysteine conformers	\\
ADIM6*	&	6	&	alkane dimers	\\
KB$^\dagger$	&	27	&	non-covalent interactions	\\
G08	&	6	&	pi-interaction of small carbon complexes 	\\
PCONF*	&	10	&	peptide conformers	\\
S66$^\dagger$	&	66	&	non-covalent interactions	\\
X40$^\dagger$	&	40	&	halogen interactions	\\
S22	&	22	&	noncovalently bound dimers 	\\
DARC	&	14	&	Diels–Alder reactions	\\
CARBHB12	&	12	&	hydrogen-bonded	\\
BHPERI	&	26	&	pericyclic reactions	\\
G21IP	&	36	&	adiabatic ionization potentials	\\
ALK8	&	8	&	alkaline compounds	\\
SCONF*	&	17	&	sugar conformers	\\
B30$^\dagger$	&	30	&	non-covalent interactions for halogen	\\
AB9	&	9	&	anomalous barrier height 	\\
SIE4x4	&	16	&	self-interaction-error 	\\
BH76	&	76	&	reaction barrier height 	\\
W4	&	140	&	atomization energies	\\
WATER27$^\dagger$	&	27	&	water interactions	\\
\hline \hline					
\end{tabular}}
\captionof{table}[Dataset description]
{Data sets used in this work. 
The asterisks indicate the data sets used for optimization of the original D3 parameters by Grimme and co-workes as described in Ref.~\cite{GAEK10}.
For the revised D3, additional 5 datasets (KB, S66, X40, B30, and WATER27) are used as a training set and marked by a dagger.
}					
\label{tbl:dbdes}					
\end{table}

\begin{table}[htb]
\centering
\resizebox{\columnwidth}{!}{\begin{tabular}{cccrrr}
\hline
$R_{\sss H-H}$    & $a$      & $b$      & training & test  & all   \\
\hline
0.9$\sim$1.5                 & 1.397 & 6.799  & 0.30 & 10.15 & 7.93 \\
2.5$\sim$3.1                 & 1.237 & -1.532 & 0.02 & 6.64  & 5.14 \\
0.9$\sim$1.5 \& 2.5$\sim$3.1 & 1.218 & 4.665  & 2.22 & 6.85  & 4.76\\ \hline
\end{tabular}}
\captionof{table}[Parameters for $H_2^+$]{
Optimized parameters in Fig.~\ref{v1-fgr:h2+} and mean absolute error (MAE) in kcal/mol. 
Note that when the toy model is applied to the HF densities, we get $a=1$ and $b=0$ regardless of the training set, and thereby both the density-driven and functional errors are eliminated. 
}
\label{tbl:h2+para1}
\end{table}

\begin{figure*}[h!]
\centering
\includegraphics[width=2.0\columnwidth]{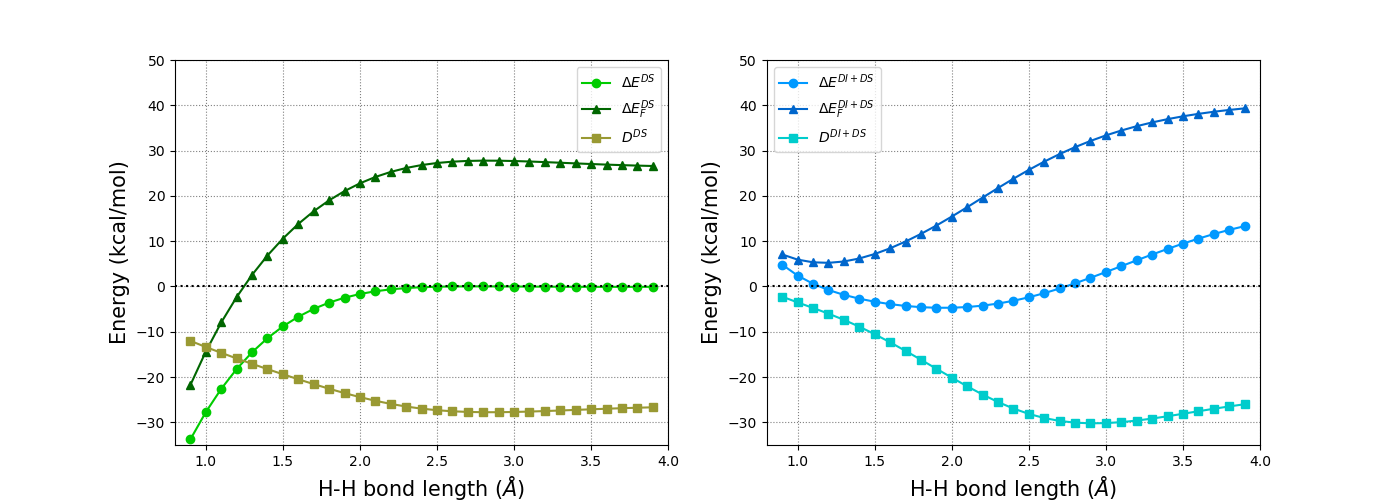}
\captionof{figure}[$\Delta E$ decomposition for $H_2^+$ curve.]{
 $D$ and $\Delta E_F$ error decomposition for H$_2^+$ dissociation curve of the empirical toy functional 
trained on the DS region (left panel) and on the combination of DS and DI regions (right panel)
when applied to H$_2^+$ along the dissociation curve . The same result, but for the functional trained on the DI region is shown in the inset of the bottom panel of of Fig.~\ref{v1-fgr:h2+}.
}
\label{fgr:divide}
\end{figure*}

\begin{figure*}[h!]
\centering
\includegraphics[width=1.0\columnwidth]{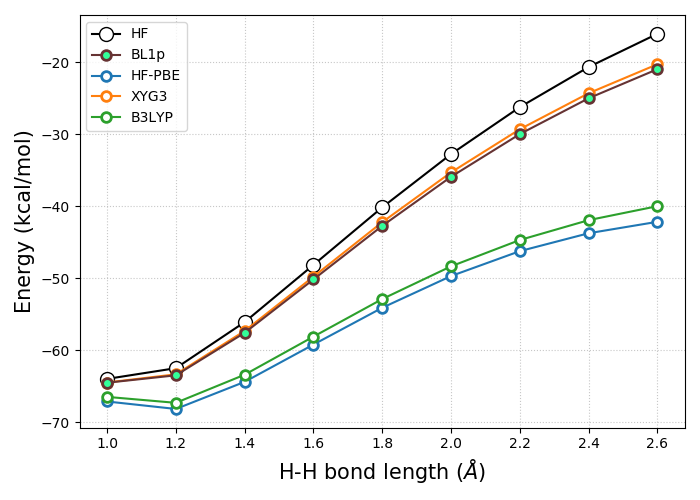}
\captionof{figure}[$H_2^+$ PES curve of BL1p]{
$H_2^+$ potential energy surface for the various methods.
}
\label{fgr:h2+pes}
\end{figure*}

\begin{figure*}[htb!]
\centering
\includegraphics[width=1.9\columnwidth]{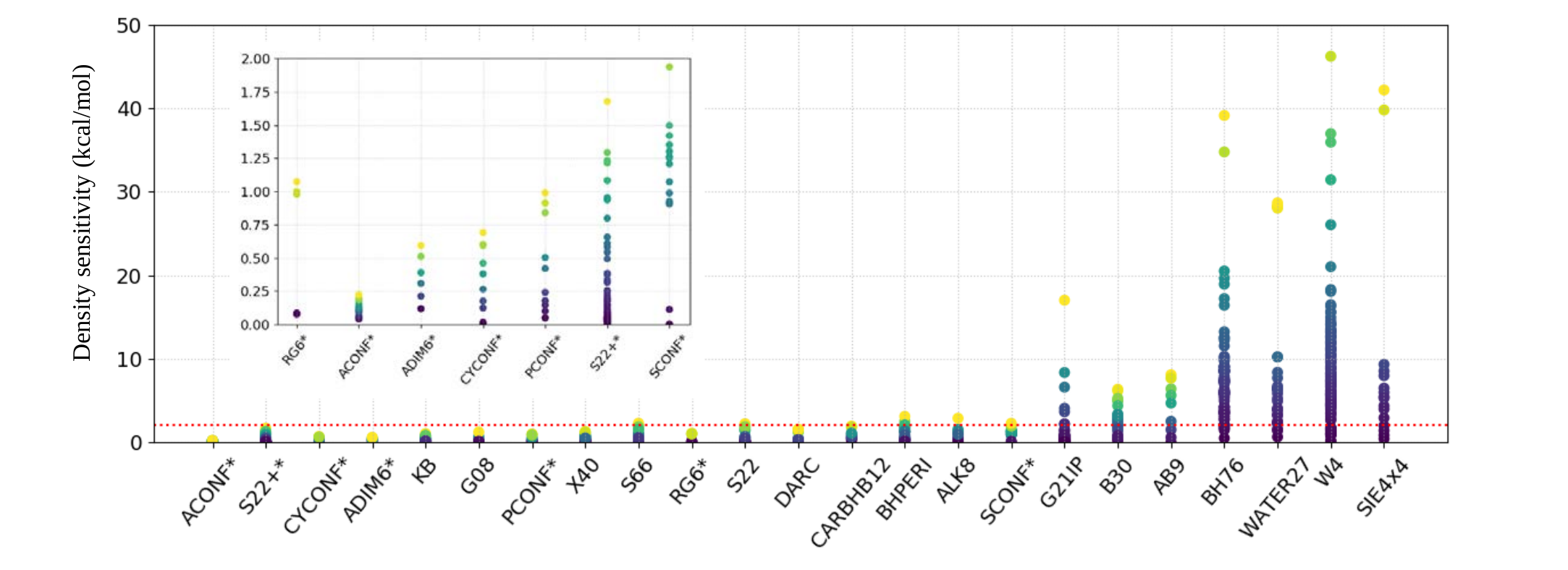}
\captionof{figure}[Densitiy sensitivity for each dataset]{
Density sensitivity ($S\pbe$) of PBE for the 23 databases in Table~\ref{tbl:dbdes}. 
The red dotted line denotes 2~kcal/mol. 
If $S\pbe$ is greater than 2~kcal/mol, it is considered density sensitive.\cite{SSB18}
As we go from the left to the right, the averaged $S\pbe$) of the databases increases. 
The databases used for the training of the Grimme's original D3 parameters are marked with an asterisk. The same databases are also shown in the inset. 
}
\label{fgr:s1}
\end{figure*}


\begin{figure}[htb]
\centering
\includegraphics[width=1.0\columnwidth]{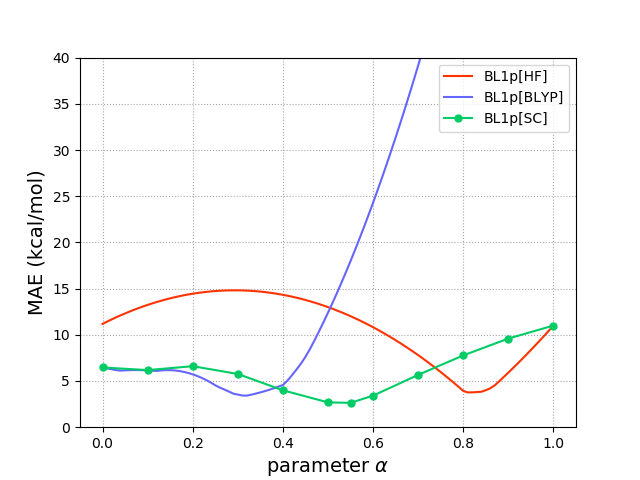}
\captionof{figure}[BL1p for AE6 dataset]{
BL1p MAE for the AE6 dataset in Ref.~\cite{STS11}.
HF, BLYP and SC density is used.
(SC denotes the self-consistent density from
$E_{XC}=\alpha E_X^{HF}+(1-\alpha)E_X^{B88}+(1-\alpha^2)E_C^{LYP}$.)
Reference values are from Ref.~\cite{PT11}.
}
\label{fgr:ae6}
\end{figure}

\begin{figure}[!htb]
\centering
\includegraphics[width=0.95\columnwidth]{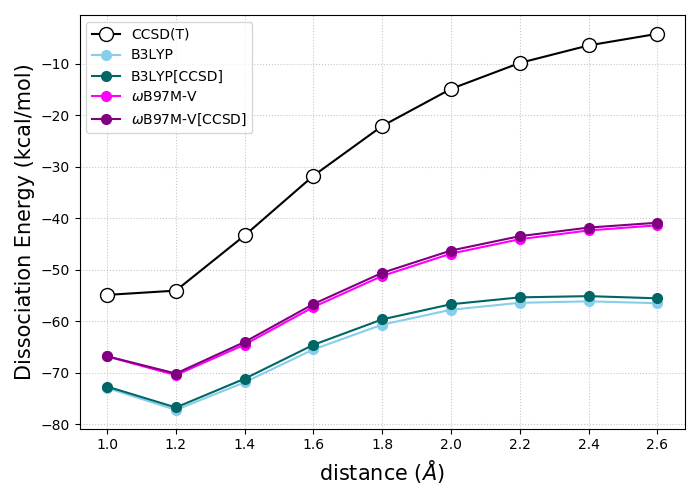}
\captionof{figure}[He$_2^+$ PES curve]{
Dissociation curve of He$_2^+$.
Note that CCSD density is used within the WY KS-inversion method
to obtain accurate KS orbitals.
Detailed information about KS-inversion can be found in the Ref.~\cite{NSSB20}.
}
\label{fgr:he2+inv}
\end{figure}

\begin{figure}[htb]
\centering
\vskip 5cm
\includegraphics[width=1.0\columnwidth]{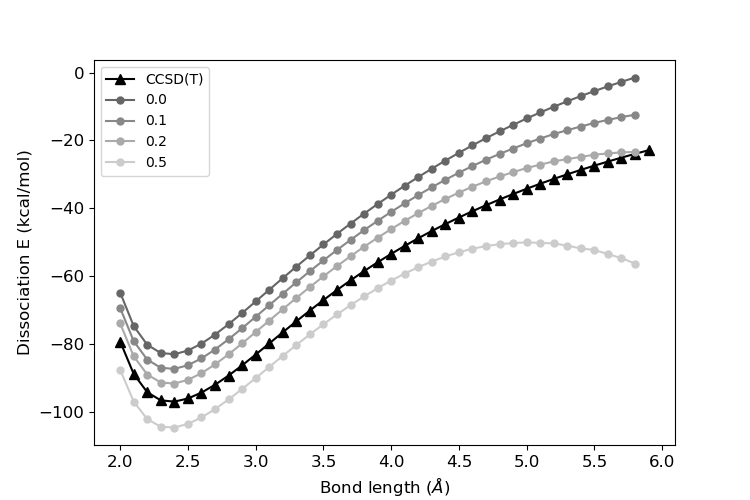}
\captionof{figure}[XYG3 for NaCl PES curve]{
Na-Cl dissociation curve for the XYG3 with different MP2 admixture. The larger the MP2 portion, the quicker the XYG3 DH bends down. 
}
\label{fgr:xyg3nacl}
\end{figure}

\clearpage
\begin{table}[]
\centering
\begin{adjustbox}{angle=90}
\begin{tabular}{lrrrrrrrrrrrr}
              & \multicolumn{1}{c}{ACONF} & \multicolumn{1}{c}{S22+} & \multicolumn{1}{c}{CYCONF} & \multicolumn{1}{c}{ADIM6} & \multicolumn{1}{c}{KB} & \multicolumn{1}{c}{G08} & \multicolumn{1}{c}{PCONF} & \multicolumn{1}{c}{X40} & \multicolumn{1}{c}{S66} & \multicolumn{1}{c}{RG6} & \multicolumn{1}{c}{S22} & \multicolumn{1}{c}{DARC} \\
BL1p          & 0.156                     & 0.263                    & 0.067                      & 0.724                     & 0.457                  & 1.061                   & 0.456                     & 0.487                   & 0.279                   & 0.350                   & 0.393                   & 1.299                    \\
BL1p*         & 0.156                     & 0.263                    & 0.067                      & 0.724                     & 0.457                  & 1.061                   & 0.456                     & 0.487                   & 0.279                   & 0.350                   & 0.393                   & 1.299                    \\
XYG3          & 0.205                     & 0.237                    & 0.179                      & 0.937                     & 0.193                  & 0.801                   & 0.090                     & 0.476                   & 0.397                   & 0.223                   & 0.293                   & 2.146                    \\
wB97M-V       & 0.056                     & 0.121                    & 0.065                      & 0.106                     & 0.161                  & 0.263                   & 0.644                     & 0.161                   & 0.092                   & 0.101                   & 0.220                   & 0.709                    \\
B3LYP         & 0.959                     & 2.231                    & 0.469                      & 5.034                     & 2.137                  & 7.467                   & 3.750                     & 1.878                   & 3.291                   & 0.494                   & 3.842                   & 15.762                   \\
B3LYP{[}HF{]} & 1.052                     & 2.435                    & 0.606                      & 5.375                     & 2.364                  & 7.771                   & 3.648                     & 2.288                   & 3.639                   & 0.482                   & 4.248                   & 16.380                   \\
M06           & 0.318                     & 0.807                    & 0.135                      & 0.275                     & 0.661                  & 1.186                   & 0.411                     & 0.601                   & 0.669                   & 0.375                   & 1.013                   & 4.642                    \\
M06-2X        & 0.240                     & 0.561                    & 0.193                      & 0.341                     & 0.452                  & 0.973                   & 1.152                     & 0.249                   & 0.253                   & 0.288                   & 0.377                   & 2.358                    \\
B2PLYP        & 0.505                     & 1.008                    & 0.230                      & 2.654                     & 0.889                  & 3.337                   & 1.575                     & 0.880                   & 1.529                   & 0.432                   & 1.694                   & 8.059                    \\
MP2           & 0.228                     & 0.204                    & 0.320                      & 0.912                     & 0.407                  & 0.936                   & 1.888                     & 0.453                   & 0.285                   & 0.335                   & 0.313                   & 1.074                   
\end{tabular}
\end{adjustbox}
\end{table}

\begin{table}[]
\begin{adjustbox}{addcode={\begin{minipage}{\width}}{\captionof{table}[MAE tables of Fig.~\ref{v1-fgr:1p}]{%
Mean absolute error (MAE, kcal/mol) of 23 databases for methods in Fig.~\ref{v1-fgr:1p}.
Note that BL1p* omits spin contamination cases for G21IP, AB9, BH76, W4, and SIE4x4.
      }
      \label{tbl:mae}
      \end{minipage}},rotate=90,center}
\begin{tabular}{lrrrrrrrrrrr}
              & \multicolumn{1}{c}{CARBHB12} & \multicolumn{1}{c}{BHPERI} & \multicolumn{1}{c}{ALK8} & \multicolumn{1}{c}{SCONF} & \multicolumn{1}{c}{G21IP} & \multicolumn{1}{c}{B30} & \multicolumn{1}{c}{AB9} & \multicolumn{1}{c}{BH76} & \multicolumn{1}{c}{WATER27} & \multicolumn{1}{c}{W4} & \multicolumn{1}{c}{SIE4x4} \\
BL1p          & 0.379                        & 3.289                      & 2.043                    & 0.168                     & 2.571                     & 0.654                   & 6.830                   & 3.751                    & 1.003                       & 4.183                  & 1.860                      \\
BL1p*         & 0.379                        & 3.289                      & 2.043                    & 0.168                     & 1.991                     & 0.654                   & 5.851                   & 2.492                    & 1.003                       & 3.077                  & 2.044                      \\
XYG3          & 0.248                        & 0.543                      & 1.295                    & 0.092                     & 1.382                     & 0.611                   & 2.785                   & 0.674                    & 1.541                       & 2.572                  & 2.411                      \\
wB97M-V       & 0.189                        & 1.145                      & 2.501                    & 0.129                     & 2.901                     & 0.450                   & 5.438                   & 1.327                    & 0.603                       & 2.113                  & 10.608                     \\
B3LYP         & 0.700                        & 4.459                      & 6.051                    & 0.771                     & 3.541                     & 1.369                   & 5.391                   & 4.336                    & 6.306                       & 4.046                  & 17.475                     \\
B3LYP{[}HF{]} & 1.076                        & 4.855                      & 6.303                    & 1.400                     & 3.928                     & 2.216                   & 7.403                   & 2.759                    & 11.051                      & 8.391                  & 12.512                     \\
M06           & 0.363                        & 2.258                      & 3.464                    & 0.381                     & 3.052                     & 1.522                   & 5.199                   & 1.831                    & 2.647                       & 2.894                  & 14.146                     \\
M06-2X        & 0.212                        & 1.351                      & 2.300                    & 0.254                     & 2.645                     & 0.931                   & 5.855                   & 1.196                    & 2.527                       & 2.955                  & 8.507                      \\
B2PLYP        & 0.385                        & 1.440                      & 3.113                    & 0.305                     & 2.294                     & 0.886                   & 3.265                   & 1.975                    & 2.613                       & 2.132                  & 9.737                      \\
MP2           & 0.358                        & 8.369                      & 3.318                    & 0.458                     & 2.600                     & 0.584                   & 6.678                   & 3.677                    & 1.238                       & 4.426                  & 2.139                     
\end{tabular}
\end{adjustbox}
\end{table}


\clearpage

\bibliographystyle{unsrt}

\label{page:end}